# Scalable Semantic Querying of Text

Xiaolan Wang[†]  Aaron Feng[‡]  Behzad Golshan[‡]  Alon Halevy[‡]
George Mihaila[‡]  Hidekazu Oiwa[‡*]  Wang-Chiew Tan[‡]

[†]University of Massachusetts  [‡]Megagon Labs
xlwang@umass.cs.edu  {aaron,behzad,alon,george,oiwa,wangchiew}@megagon.ai

## ABSTRACT

We present the KOKO system that takes declarative information extraction to a new level by incorporating advances in natural language processing techniques in its extraction language. KOKO is novel in that its extraction language simultaneously supports conditions on the surface of the text and on the structure of the dependency parse tree of sentences, thereby allowing for more refined extractions. KOKO also supports conditions that are forgiving to linguistic variation of expressing concepts and allows to aggregate evidence from the entire document in order to filter extractions.

To scale up, KOKO exploits a multi-indexing scheme and heuristics for efficient extractions. We extensively evaluate KOKO over publicly available text corpora. We show that KOKO indices take up the smallest amount of space, are notably faster and more effective than a number of prior indexing schemes. Finally, we demonstrate KOKO's scaleup on a corpus of 5 million Wikipedia articles.

## 1. INTRODUCTION

Information extraction is the task of extracting structured data from text. Information extraction systems typically try to extract named entities (e.g., business or people names) or facts in the form of triples (e.g., (FRANCE, CAPITAL, PARIS)). Systems for information extraction fall into two main categories. The first category is machine-learning based systems, where a significant amount of training data is required to train a good model for performing a specific extraction task, such as extracting company names or company CEOs. The second category consists of rule-based systems which employ a declarative language for specifying patterns that identify the desired data in the text. Unlike machine-learning based models which tend to be opaque, rules/queries produced by rule-based systems can be analyzed and therefore the results they extract tend to be explainable [13]. For many machine learning models, preparing an annotated corpus is typically the most expensive and time consuming process [34]. Because rule-based systems do not require training data, they are applicable more broadly and can also be used as a tool for obtaining an initial annotated corpus for distant supervision. Furthermore, as we shall describe, KOKO is more than an extraction system. It is a query language that can also be used to obtain insights to a corpus.

We describe the KOKO system for extracting structured information (a relation of text and different entity types) from text. KOKO takes declarative information extraction to a new level by incorporating advances in natural language processing techniques into the extraction language. KOKO supports the new features with efficient and scalable database-style querying. We illustrate the features of KOKO with examples.

The main construct used in extraction languages and supported also in KOKO is regular expressions over the surface text of a sentence with conditions on the POS (part of speech) tags [33] of the words. For example, suppose we want to extract foods that were mentioned as delicious in text. For the sentence "*I ate delicious cheese cake*", it would suffice to specify an extraction pattern that looks for the word "*delicious*" preceding a noun that is known to be in the category of foods. However, specifying a regular expression that would extract correctly from the sentence "*I ate a delicious and salty pie with peanuts*" is trickier because the word "*delicious*" does not immediately precede the word "*pie*", and it also precedes the word "*peanuts*" which were not deemed delicious. In the sentence "*I ate a chocolate ice cream, which was delicious, and also ate a pie*" the word "*delicious*" comes *after* "*ice cream*" which makes the extraction even more challenging.

To address such challenges, KOKO supports patterns that exploit the semantic structure of text sentences. This structure is represented as *dependency parse trees* (or *dependency trees* in short). In recent years, it has become possible to efficiently create dependency trees for large corpora [11, 22, 40]. The dependency tree of our last example is shown in Figure 1. The dependency tree shows that the word "*delicious*" is in the subtree of the direct object of the verb "*ate*", which is the "*chocolate ice cream*". As we explain later, this intuitively means that "*delicious*" refers to "*chocolate ice cream*". Hence, a pattern over the dependency tree that looks for the word "*delicious*" in the subtree of a noun in the food category could provide the correct extractions.

While the idea of specifying extractions by leveraging dependency trees has been explored in the past [38, 44], KOKO is novel in that it provides a single declarative language that combines the surface-level patterns with the tree patterns and uses novel indexing techniques to scale to large corpora.

The final element of the KOKO language allows for extractions that accommodate variation in linguistic expression and aggregation of evidence. Consider the task of extracting cafe names from blog posts. Cafe names vary quite a bit, and it is nearly impossible to write rules that would extract them accurately. On the other hand, it may be possible to combine evidence from multiple mentions in the text to extract cafe names with high confidence. For example, if we see that an entity *employs baristas* and *serves espresso*, we might infer that it is a cafe. However, there are many linguistic variations on how these properties are expressed, such as *serves up delicious cappuccinos*, or *hired the star barista*. KOKO includes a semantic similarity operator that retrieves phrases that are linguistically similar to the one specified in the rule. Semantic similarity can be determined using paraphrase-based word embeddings. KOKO attaches a confidence value to the phrases matched by the similarity operator, and these confidence values can be aggregated from multiple pieces of evidence in a document. For example, if we see that an entity *serves great macchiatos* and *recently hired a barista* we may have enough evidence that it is a cafe.

---

[*]The author is now at Google Japan.

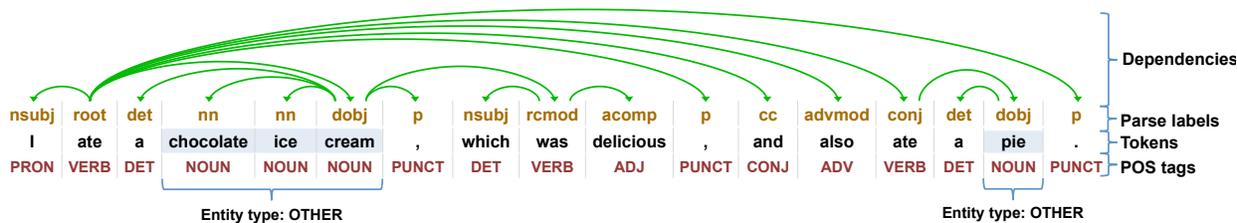

Figure 1: A sentence and its dependency tree annotated with parse labels [28], tokens, POS tags [33], and entity types. This dependency tree is generated from Google Cloud NL API [22].

In summary, our contributions are the following.

- We present the KOKO query/extraction language that combines patterns on the text surface, patterns on the semantic structure of sentences, a phrase similarity operator that accounts for linguistic variation in text, and the ability to aggregate evidence from different parts of the text.
- We describe heuristics and a novel multi-indexing scheme for quickly pruning irrelevant sentences therefore enabling KOKO to support extraction at scale. Our multi-indexing scheme comprises of a word index, and indices of different metadata of dependency trees.
- We have open-sourced part of KOKO's code and we plan to make the full KOKO code available.
- We demonstrate KOKOs efficiency and effectiveness through a suite of experiments on public text corpora. We show that KOKO indices take up the least amount of space, are at least 7 times faster and 1.7 times more effective than some prior indexing techniques. Moreover, through its collective evidence-based extraction method, KOKO is up to 3 times better in terms of precision and recall than other techniques for extracting cafe names. Finally, we demonstrate that KOKO scales up with experiments on 5 million Wikipedia articles and KOKO's performance is linear in the number of articles.

**Outline of paper** We overview the KOKO language in Section 2 before we describe the indices we built in Section 3 and show how we evaluate a KOKO query in Section 4. We describe related work in Section 5, demonstrate the efficiency and scalability of KOKO in Section 6 before we conclude and discuss future work in Section 7.

## 2. THE KOKO LANGUAGE

The KOKO language supports three kinds of constructs: (1) conditions on the surface text with regular expressions, (2) conditions on the hierarchical structure of the dependency tree, and (3) linguistic similarity conditions whose results can be aggregated across the entire document. We describe the first two constructs in Section 2.1 and the third in Section 2.2.

**Preprocessing the input:** The input to a KOKO query is a text document. We first process the document with a natural language parser (e.g., spaCy [40] or Google NL API [22]). The preprocessing transforms the document into a sequence of sentences, each of which consists of a sequence of *tokens*. Each token carries a number of annotations, such as the POS tag, parse label, and a reference to the parent in the dependency tree. We refer to a sequence of consecutive tokens in a sentence as a *span*.

### 2.1 Surface and hierarchy conditions

To support conditions on the surface text and the dependency tree, variables in a KOKO expression can be bound to two kinds of terms. *Node terms* refer to the nodes in the dependency tree of the sentence, and *span terms* refer to spans of text. Given a node $x$, $x$.subtree refers to the span that includes all the words in the subtree of $x$. Given a node $x$, we also use $x$ to refer to the span that includes only the text of the word $x$.

The output of a KOKO expression is a bag of tuples. The values in the tuples can be either nodes or spans. The tuples to be returned are defined by the extract clauses (that correspond roughly to the select and where clauses in SQL). We now describe how to construct terms and specify conditions on terms and variables.

**Node terms:** Node terms are defined using XPath [46]-like syntax. A path is defined with the "/" (child) or "//" (descendant) axes and each axis is followed by a label (a parse label, POS tag, token, wildcard (*), or an already defined node variable). The expression $a$ = //verb binds the node variable $a$ to verb nodes which may occur arbitrarily deep under the root node. The node variable $b$ = $a$/dobj binds dobj nodes directly under nodes of $a$ to $b$. The path $b$//"delicious" binds a node with token "*delicious*" that may occur arbitrarily deep under $b$ to variable $c$. Observe that the condition touches upon different annotations: verb is a POS tag, dobj is a parse label, while "*delicious*" is a word.

Each label can be associated with conditions, which are specified in [...], such as a regular expression [@regex = ⟨regular expression⟩] or [@pos="noun"], which states that the POS tag of the current node must be equal to noun. Hence, writing /root//noun is the same as writing /root//*[@pos="noun"], where ∗ denotes an arbitrary label and the POS tag of the label must be noun. Multiple conditions stated within [...] are separated by ",". For example, [@pos="noun", etype="Person"] states that the POS tag is noun and the entity type is Person.

**Span terms:** A span term $x$ is constructed with the syntax $x$ = $\langle atom \rangle_1 + \ldots + \langle atom \rangle_k$, where $k \geq 1$ and $\langle atom \rangle_i$ is one of the following: a path expression as described above, a node variable, a sequence of tokens, $x$.subtree, or an elastic span ∧ (which denotes zero or more tokens) or with conditions ∧[...]. For ∧, KOKO also allows the specification of regular expression over the span or the minimum and/or maximum number of tokens for the span. For example, $x$= //verb + $a$ + $b$.subtree + ∧[etype="Entity"] defines $x$ to bind to a span that must begin with //verb, followed immediately by the span of $a$, the span given by $b$.subtree, and the span that defines an entity in this order. A span term has type Str (String). The output of a Koko query also serializes all entity types into strings.

**The Extract clause:** The extract clause is where variables are defined. The defined variables also need to satisfy certain conditions: typed conditions, hierarchical structural conditions over the dependency tree and/or horizontally over the sequence of tokens. Users can specify constraints among the variables outside a block using the in or eq constructs. For example, "$x$ in $y$" requires that the tokens of $x$ are among the tokens of $y$, while "$x$ eq $y$" requires that the two spans given by $x$ and $y$ are identical.

EXAMPLE 2.1. The small example query below extracts pairs $(e,d)$ where $e$ is an entity type and $d$ is a string type. The variables $a$, $b$, $c$, and $d$ are defined within the *block* /ROOT:{ ...} where the paths are defined w.r.t. the root of the dependency tree. The variables $a$, $b$, and $c$ are node terms while $d$ and $e$ (defined in the first line) are span terms. Outside the block, "$(b)$ in $(e)$" is a constraint between $b$ and $e$ which asserts that the dobj token must be among the tokens that make up entity $e$.

> extract $e$:Entity, $d$:Str from input.txt if
> (/ROOT:{
>     $a$ = //verb,
>     $b$ = a/dobj,
>     $c$ = b//"delicious",
>     $d$ = ($b$.subtree)
> } $(b)$ in $(e)$)

For the sentence in Figure 1, there is only one possible set of bindings for the variables in this query: $a$ = "*ate*", $b$ = "*cream*", $c$ = "*delicious*", $d$ = "*a chocolate ice cream, which was delicious*", and $e$ = "*chocolate ice cream*". The query returns the pair $(e,d)$. □

## 2.2 Similarity and aggregation conditions

We specify additional constraints on the variables in the satisfying clause. Some of these constraints are boolean and some are approximate and return a confidence value. We use an example of extracting cafe names from authoritative blog posts in this section.

The satisfying clause contains a disjunction of *boolean* or *descriptor conditions*. Boolean conditions, which include conditions specified with regular expressions, evaluate to true or false. For example, $x$ ", *a cafe*" requires that $x$ is immediately followed by the string ", *a cafe*", and is a sufficient condition for determining that $x$ is the name of a cafe. Other types of conditions using matches or contains are also allowed (see Section 4.4.1).

**Descriptors** A descriptor condition evaluates to a confidence value. There are two types of descriptor conditions. The first is of the form $x$ similarTo ⟨descriptor⟩ which returns how similar is $x$ to ⟨descriptor⟩. The second is of the form $x$ [[*descriptor*]] (or [[*descriptor*]] $x$) and returns how similar *descriptor* is to the span after $x$ (or before $x$). Note that the distance between $x$ and the terms similar to *descriptor* affects the confidence returned for the similarity.

Every condition is associated with a weight between 0 and 1, which specifies how much emphasis to place on the condition when a match occurs. Upon receiving the query, each descriptor is expanded to words semantically close to it using a paraphrase embedding model. In the above example, entities that serve coffee could also be considered cafes. However, instead of specifying all different ways to capture the meaning that a cafe sells coffee, the user can specify the condition ($x$ [["serves coffee"]]). The expansion will yield similar phrases such as "*sells espresso*" and "*sells coffee*".

EXAMPLE 2.2. Here we present an example of how we are able distinguish two syntatically equivalent sentences through our similarity description "similarTo".

*S1: cities in asian countries such as china and japan.*
*S2: cities in asian countries such as beijing and tokyo.*

The above sentences have the same syntax, but we are able to distinguish them through the following KOKO queries ("GPE" refers to location entities).

Q1:
> extract $a$:GPE from "input.txt" if ()
> satisfying $a$
>     ($a$ SimilarTo "city" {*1.0*})

Q2:
> extract $a$:GPE from "input.txt" if ()
> satisfying $a$
>     ($a$ SimilarTo "country" {*1.0*})

The first query Q1 retrieves cities while the second query retrieves countries. If we execute the above two KOKO queries on the two sentences, we obtain:

|    | S1 | S2 |
|----|----|----|
| Q1 | NA | Tokyo, 0.4092163796959<br>Beijing, 0.3577236986 |
| Q2 | China, 0.512509382446<br>Japan, 0.45709525868 | NA |

It is also possible to write a single Koko query to retrieve the correct is-a relation over the two sentences so that the result will be (countries, china), (countries, japan), (cities, beijing), (cities, tokyo). We omit the query here. □

Although the expansion is not always perfect, descriptors enable users to be agnostic to linguistic variations to a large extent.[1]

**Aggregation** For every sentence where the extract clause is satisfied, KOKO will search the text corpus to compute a score for every satisfying clause. It does so by computing, for every sentence, a score that reflects the degree of match according to the conditions and weights in the satisfying clause and then aggregating the scores of the sentences. For every variable, if the aggregated score of the satisfying clause for that variable from the collective evidence passes the threshold stated, then the result is returned. We give a detailed description of the aggregation semantics in Section 4.4.

EXAMPLE 2.3. The query below has an elaborate satisfying clause (with empty extract clause). The intent of the query is to extract cafe names and considers all entities as candidate cafe names. However, only those that pass the satisfying $x$ clause will be returned as answers.

The first two boolean conditions checks whether the name of the entity contains "*Cafe*" or "*Roasters*". It also looks for evidence in input.txt that the name is followed by the string ", *a cafe*". In addition, it will search the text for evidence that the name is followed by a phrase that is similar to "*serves coffee*" or "*employs baristas*".

> extract $x$:Entity from "input.txt" if ()
> satisfying $x$
>     (str($x$) contains "Cafe" {*1*}) or
>     (str($x$) contains "Roasters" {*1*}) or
>     ($x$ ", a cafe" {*1*}) or
>     ($x$ [["serves coffee"]] {*0.5*}) or
>     ($x$ [["employs baristas"]] {*0.5*})
> with threshold 0.8
> excluding (str($x$) matches "[Ll]a Marzocco")

The first 3 conditions have equal weights of 1, and the last two only 0.5. When we run the above query over authoritative coffee sites where new cafes are described, it is highly unlikely that we will find exact matches of the phrase "*serves coffee*", which is why the descriptor conditions play an important role. The excluding condition ensures that $x$ does not match the string "*La (or la) Marzocco*", which refers to an espresso machine manufacturer. □

To summarize, a basic KOKO query has the form, where there can be up to one satisfying clause for each output variable.

> extract ⟨*output tuple*⟩ from ⟨*input.txt*⟩ if
>   ⟨*variable declarations, conditions, and constraints*⟩
> [satisfying⟨*output variable*⟩
>   ⟨*conditions for aggregating evidence*⟩
> with threshold $\alpha$]
> [excluding ⟨*conditions*⟩]

---

[1] One can also supply a dictionary of different types of coffee to KOKO to guide the expansion.

## 3. INDEXING THE TEXT

To process queries efficiently, it is crucial that KOKO maintain several indexes on the text. We describe the indices below and our experiments will demonstrate that they enable us to speed up query processing by at least a factor of 7. The indices are built offline or created when the input text is first read. Indices can be persisted for subsequent use.

KOKO has two types of indices: *inverted index* and *hierarchy index*. We create *inverted indices* for words and entities and *hierarchy indices* for parse labels and POS tags. Unlike indices of [7, 20] our hierarchy index is a compressed representation over dependency structure for parse labels and POS tags. By merging identical nodes, our hierarchy index reduces more than 99.7% of the nodes for both parse labels and POS tags. Hierarchy indices are therefore highly space efficient and enables fast searching.

EXAMPLE 3.1. *The sentence in Figure 1 has the following sentence ids and token ids, which are written above the tokens.*

sid: 0  $\overset{0}{I}\ \overset{1}{ate}\ \overset{2}{a}\ \overset{3}{chocolate}\ \overset{4}{ice}\ \overset{5}{cream}\ \overset{6}{,}\ \overset{7}{which}\ \overset{8}{was}\ \overset{9}{delicious}\ \overset{10}{,}\ \overset{11}{and}\ \overset{12}{also}\ \overset{13}{ate}\ \overset{14}{a}\ \overset{15}{pie}\ \overset{16}{.}$

Another example sentence is shown below. The sentence id and token ids are depicted after the sentence.

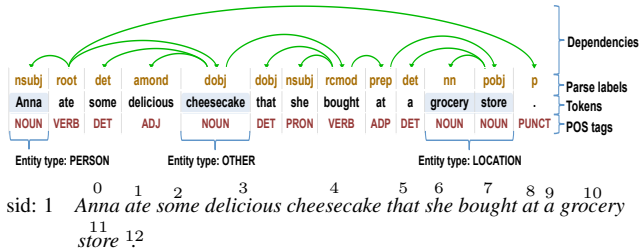

sid: 1  $\overset{0}{Anna}\ \overset{1}{ate}\ \overset{2}{some}\ \overset{3}{delicious}\ \overset{4}{cheesecake}\ \overset{5}{that}\ \overset{6}{she}\ \overset{7}{bought}\ \overset{8}{at}\ \overset{9}{a}\ \overset{10}{grocery}\ \overset{11}{store}\ \overset{12}{.}$

### 3.1 Inverted indices

KOKO has two inverted indices: *word index* and *entity index*. One can also have more refined entity indices, such as Person index, Organization index and so on but we do not describe them here.

A *word index* maps words to sentences that contain them along with relevant metadata. Specifically, every word in the index points to a list of quintuples $(x,y,u\text{-}v,d)$: the sentence id ($x$) and token id ($y$) of the word of interest in the sentence, the first ($u$) and last ($v$) token id of the subtree rooted at the current token based on the dependency tree, and the depth ($d$) of the token in the dependency tree. An *entity index* is defined similarly. Here, we track the sentence id and the leftmost and rightmost token ids, corresponding to the span of the entity name in the sentence. Like the word index, the entity index facilitates quick access to the locations of all sentences where this entity occurs.

EXAMPLE 3.2. *Part of the word index and entity index based on the sentences in Figure 1 and Example 3.1 are shown below.*

| word | list of quintuples |
|---|---|
| I | (0,0,0-0,1) |
| ate | (1,1,0-12,0), (0,1,0-16,0) |
| delicious | (1,3,3-3,2), (0,9,9-9,3) |
| cream | (0,5,2-9,1) |

| entity | list of triples |
|---|---|
| cheesecake | (1,4-4) |
| grocery store | (1,10-11) |
| choc. ice cream | (0,3-5) |

Since "*ate*" (sid=1, tid=1) is the root of the dependency tree of the second sentence, the left and right reachable token ids correspond to the first and last token of the sentence, which are 0 and 12 respectively, and the depth of this token is 0. The word index facilitates quick access to the locations of all sentences where this word occurs and allows one to check for parent-child relationship. For example, if two tokens $t_p$ and $t_c$ satisfy the following condition: $t_p.x = t_c.x \land t_p.u \leq t_c.u \land t_p.v \geq t_c.v \land t_p.d = t_c.d + 1$, then we know that $t_p$ is the parent of $t_c$. Existing indexing techniques such as [7, 20] for the *constituency-based parse trees* also contain a similar set of information to index parse labels. □

### 3.2 Hierarchy indices

A hierarchy index is a compact representation of all dependency trees, which provides fast access to the dependency structure of all sentences. A hierarchy index is constructed by merging identical nodes of dependency trees of all sentences together. Starting at the root of all dependency trees, children nodes with the same label are merged, and then for each child node, all children nodes with the same labels are merged and so on. Hence, by construction, every node in a hierarchy index has a set children nodes with distinct labels. Consequently, every node of the index can be identified through a unique path given by the sequence of labels from the root node to that node. Every node is annotated with a *posting list*, which tracks tokens of sentences that have the given path.

KOKO has two hierarchy indices: the *hierarchy index for parse labels (PL index)* and the *hierarchy index for POS tags (POS index)*, which are constructed by merging dependency trees based on parse labels and, respectively, POS tags.

EXAMPLE 3.3. *The table below shows part of the posting list for the PL index where each node is identified by its unique path from the root to that node. A posting list is a list of quintuples, identical to what was stored in the word index. For readability, we have also annotated each quintuple with the word it represents.*

Observe that both nn nodes directly under the dobj node of the sentence in Figure 1 have been merged. Hence the posting list associated with /root/dobj/nn contains both "*chocolate*" and "*ice*". Since the root of every dependency tree has parse label *root*, there is a single PL index for all dependency trees based on parse labels starting from the *root* label. However, this is not the case for POS tags. We assume there is a dummy node with the same dummy label above the root of every dependency tree in this case to merge all dependency trees based on POS tags into one POS index.

| node | posting list |
|---|---|
| /root | *ate*(1,1,0-12,0), *ate*(0,1,0-16,0) |
| /root/nsubj | *Anna*(1,0,0-0,1), *I*(0,0,0-0,1) |
| /root/dobj | *cheesecake*(1,4,2-11,1), *cream*(0,5,2-9,1) |
| /root/dobj/det | *some*(1,2,2-2,2), *a*(0,2,2-2,2) |
| /root/dobj/amod | *delicious*(1,3,3-3,2) |
| /root/dobj/nn | *chocolate*(0,3,3-3,2), *ice*(0,4,4-4,2) |

A hierarchy index is similar to a strong dataguide [21], which is a structural summary of semistructured data. As already pointed out in [21], a strong dataguide can be used as a path index for fast access to elements of that path. To the best of our knowledge, however, this is the first use of different hierarchy indices as a compact representation of dependency trees to speed up query processing over text data. It is also the first application that involves simultaneous access to hierarchy indices and inverted indices to process queries that involve different types of conditions.

## 4. EVALUATING KOKO QUERIES

The basic workflow of the KOKO system is depicted in Figure 2. The input to the KOKO engine is a query and a text corpus. Initially, we parse the text corpus to obtain the dependency trees and create the PL index, POS index, word index, and entity index. We use a relational database management system (PostgresSQL), to store the parsed text and the constructed indices. The inverted indices have flat structure and therefore, can be directly stored in relational tables. We use Closure tables [25] to represent the hierarchy index which we will describe in more detail in Section 6.

Next, we describe the workflow of KOKO system for evaluating a given query, and in particular, how the indices and heuristics are exploited to achieve efficiency. There are 4 main processing steps:

1. *Normalize query.* The path expressions in the extract clause are first normalized and conditions among variables are explicitly stated in preparation for subsequent steps.

2. *Decompose paths and lookup indices.* Every path expression from above is decomposed into one or more paths so that each decomposed path can be used to access an index. The results from all accesses to indices are then joined, as needed, to obtain a candidate set of sentences that should be considered next.

3. *Generate skip plan.* This module applies a heuristic on "horizontal" conditions to identify a set of variables whose evaluation can be first skipped. The bindings for the skipped variables are then derived and constraints are checked.

4. *Aggregate.* The conditions of satisfying clauses are evaluated across the text to obtain the final result.

## 4.1 Normalize query

The path expressions in the extract clause are first expanded into their absolute form and constraints among variables are explicitly stated. For example, the path expressions of node variables $a, b, c, d$ in Example 2.1 will be expanded as follows:

$a$ = //verb,  $b = a$/dobj $\longmapsto b$ = //verb/dobj,  $d = b$.subtree
$c = b$/"delicious" $\longmapsto c$ = //verb/dobj/"delicious"

In addition, two constraints ($a$ parentOf $b$) and ($b$ ancestorOf $c$) will be added.

EXAMPLE 4.1. To exemplify further, consider the KOKO query, the corresponding variables and path expressions, and the derived constraints shown below:

**Query:**  extract $a$:Str,$b$:Str,$c$:Str from input.txt if (
/ROOT:{
$a$ = Entity, $b$ = //verb[text="ate"],
$c = b$/dobj, $d = c$//"delicious",
$e = a + \wedge + b + \wedge + c$ })

**Variables & path expressions:**
$a$=Entity, $b$=//verb[text="ate"]
$c$=//verb[text="ate"]/dobj
$d$=//verb[text="ate"]/dobj//"delicious"
$v_1 = \wedge$, $v_2 = \wedge$

**Constraints:**
$b$ parentOf $c$, $c$ ancestorOf $d$,
$a$ leftOf $v_1$, $v_1$ leftOf $b$,
$b$ leftOf $v_2$, $v_2$ leftOf $c$.

The last condition in the query states that the span consisting of $a$, $b$, and $c$ must occur in this order in the sentence and there can be zero or more tokens between $a$ and $b$ and between $b$ and $c$. After normalizing the query, variables $v_1$ and $v_2$ are automatically assigned for tracking the respective $\wedge$, and associated adjacency constraints are also derived. □

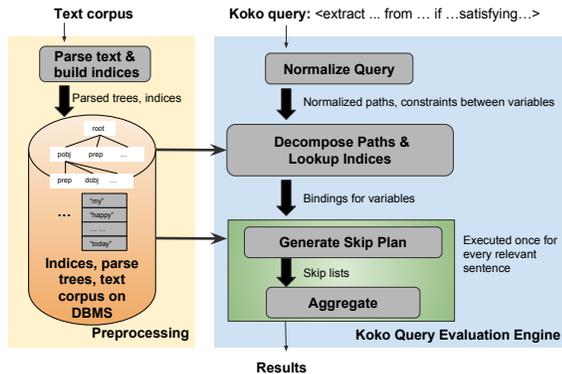

Figure 2: Overview of KOKO system.

**Algorithm 1:** Decompose paths and lookup indices

**Input** : normalized query, indices
**Output**: candidate bindings for each variable

*bindings* = empty;
**foreach** *variable x defined as entities* **do**
  *bindings[x]* = union of posting lists from accessing entity index;
*dom_paths* = set of dominant paths in query;
*pbindings* = empty;
**foreach** *p in dom_paths* **do**
  Decompose $p$ into parse label path $p_1$, POS tag path $p_2$, and word path $w$ if possible (details in text).
  $P_1$ = lookup PL index with $p_1$; union the resulting posting lists;
  $P_2$ = lookup POS index with $p_2$; union the resulting posting lists;
  $Q$ = lookup word index with words in $w$ and join results;
  *pbindings[p]* = join $P_1$, $P_2$, and $Q$;
**foreach** *variable x defined as a path* **do**
  Let $p$ be the dominant path of $x$;
  *bindings[x]* = *pbindings[p]*;
**return** *bindings*;

## 4.2 Decompose path and lookup indices

This module (DPLI for short) decomposes paths from the previous step to access the indices (see Algorithm 1). DPLI first looks for variables defined as entities and returns the posting list by looking up the entity index. For example, variable $a$ from Example 4.1 will bind to the union of all posting lists of all entities in the index.

### 4.2.1 Decompose paths

Next, among the variables defined by paths, DPLI looks for paths that are *not dominated* by other paths and only decomposes undominated paths to access the indices.

A path $p$ is *dominated by* a path $q$ if (1) $p$ without conditions is a prefix of $q$ without conditions, and (2) every condition of a label in $p$ is identical to the condition of the corresponding label in $q$ modulo order of conjunction. A path is *dominant* if it is not dominated by any other path. In Example 4.1, $d$ is the only dominant path.

We decompose only dominant paths for index lookup. This is because by definition, for any given path $r$ along a dominant path $p$, the nodes with path $r$ which have descendants among the nodes of $p$ are a subset of the nodes of $r$. Hence, the nodes of the dominant path can be used to find the ancestor nodes that satisfy the conditions for all of its dominated paths.

For every dominant path, we decompose it into one or more paths so that each decomposed path can be used to access a specific type of index. Every dominant path $p$ is decomposed into up to 3 paths: (1) a *parse label path*, (2) a *POS tag path*, and (3) a *word path*. Paths from (1), (2), and (3) are used to access the PL index, the POS index, and the word index respectively before the results are "joined" to obtain a set of candidate bindings for further processing.

**Parse label, POS tag, and word paths** Assume a path $p$ has the form $\#l_1\#...\#l_m$, where "#" denotes the axis "/" or "//". A *parse label path* $p_1$ (resp. *POS tag path* $p_2$; *word path* $w$) is extracted from $p$ by replacing every $l_i$, where $1 \leq i \leq m$, with $*$ if $l_i$ is *not* a parse label (resp. not a POS tag; not a word).

EXAMPLE 4.2. Consider the normalized variable path
d = //verb[text="ate"]/dobj//"delicious" in Example 4.1, we derive the following parse label, POS tag, and word paths:

Parse label path ($p_1$):    //*/dobj//*
POS tag path ($p_2$):        //verb/*//*
Word path ($w$):             //"ate"/*//"delicious"           □

### 4.2.2 Lookup indices and join

**Lookup PL/POS index and union posting list** With the parse label path $p_1$ and POS tag path $p_2$, we lookup the PL index and POS tag index with $p_1$ and $p_2$ respectively. The resulting posting lists are unioned in each case and we denote the unioned results by $P_1$ and $P_2$ respectively. For each word along the word path $w$ (from left to right), we access the word index and *join* the posting lists obtained as we traverse the words on the word path to ensure the ancestor-descendant relationships among the words are enforced.

EXAMPLE 4.3. Continuing with Example 4.2, the PL index of Section 3.2 is accessed with $p_1$ to obtain the following set $P_1$ of posting lists (only partially shown). The result $P_2$ of accessing the POS index is not shown.

| /root/dobj/det | *some*(1,2,2-2,2), *a*(0,2,2-2,2) |
| /root/dobj/amod | *delicious*(1,3,3-3,2) |
| /root/dobj/nn | *chocolate*(0,3,3-3,2), *ice*(0,4,4-4,2) |

**Lookup word index and join posting list** This description refers to the computation of $Q$ in the 2nd for loop of Algorithm 1. We *join* $Q_1$ with $Q_2$ as follows. For every $(x_1, y_1, u_1\text{-}v_l, l_1) \in Q_1$ and $(x_2, y_2, u_2\text{-}v_2, l_2) \in Q_2$, we return $(x_2, y_2, u_2\text{-}v_2, l_2)$ as the *join* of $Q_1$ with $Q_2$ if $x_1 = x_2$, $u_1 \leq u_2$, $v_1 \geq v_2$, and $l_2 \geq l_1 + 2$. The condition $x_1 = x_2$ ensures that the words are from the same sentence. The second and third conditions ensure that the former word is an ancestor of the latter in the dependency tree.

If there are more words along the word path, the current posting list will be joined with the next posting list, which is obtained by accessing the word index with the next word along the word path. This process repeats until every word along the path has been considered. The final result is a set of quintuples that respect the ancestor-descendant relationship as specified in the word path.

EXAMPLE 4.4. For the word path //"ate"/*//"delicious", the word index is first accessed with "*ate*" (the leftmost word of $w$) to obtain $Q_1 = \{(1,1,0\text{-}12,0), (0,1,0\text{-}16,0)\}$ and then accessed with "*delicious*" to obtain $Q_2 = \{(1,3,3\text{-}3,2), (0,9,9\text{-}9,3)\}$. After this, we join $Q_1$ with $Q_2$. The word path allows us to conclude that "delicious" is at least at depth 2 higher than "ate", and so we have the condition $l_2 \geq l_1 + 2$. The join of $Q_1$ and $Q_2$ for this word path is $\{(1,3,3\text{-}3,2), (0,9,9\text{-}9,3)\}$. This posting list will later be joined with the posting lists from PL index and POS tag index respectively. □

**Join of posting lists from all indices** This description refers to the computation of *pbindings* in the 2nd for loop of Algorithm 1. After we obtain $P_1$, $P_2$, and $Q$, we first join $P_1$ with $P_2$ to obtain a single list $P$, which is computed as follows. For every $(x_1, y_1, u_1\text{-}v_1, l_1) \in P_1$ and $(x_2, y_2, u_2\text{-}v_2, l_2) \in P_2$, the quintuple $(x_2, y_2, u_2\text{-}v_2)$ is returned if $x_1 = x_2$ and $y_1 = y_2$. In other words, we look for quintuples that refer to the same token in this join.

Finally, we join $P$ and $Q$. The join of $P$ and $Q$ depends on two cases. In our running example, the last element of the word path is a word token (i.e., "*delicious*"). So we look for quintuples in $P$ and $Q$ that refer to the same token of the same sentence. In other words, we return $(x_1, y_1, u_1\text{-}v_1, l_1)$ if $x_1 = x_2$ and $y_1 = y_2$. If the last element is not a word token (e.g., //"ate"/*/*) then we ensure that the quintuple of $Q$ an ancestor of the quintuple of $P$ with the appropriate depth requirements and the quintuple of $P$ is returned as a binding for the path $p$. The join condition for this example is expressed as $x_1 = x_2$ and $u_2 \leq u_1$ and $v_2 \geq v_1$ and $l_2 + 2 = l_1$ and we return $(x_1, y_1, u_1\text{-}v_1, l_1)$ if the conditions are satisfied.

After the join is completed, the final result returned defines the candidate bindings for each variable according the bindings of its dominant path. (See last for loop of Algorithm 1.)

**Discussion** One case that is not described in the algorithm is what happens when the path used to access an index does not exist in the index. If this happens, the evaluation immediately ceases and returns with an empty answer.

Otherwise, the final join result is a set of quintuples that refers to tokens whose paths satisfy every decomposed path of $p$. Observe that if a token of a sentence has a path $p'$ in the dependency tree that satisfies $p$, then $p'$ must satisfy every decomposed path of $p$. However, the converse may not always hold. In general, there may be tokens of sentences whose path $p'$ satisfy every decomposed path of $p$ but $p'$ does not satisfy $p$. Hence, the resulting set of bindings for each path $p$ that are returned are complete but it may still contain results that are not part of the answers. The Generate Skip Plan will include a validation step to ensure that the wrong answers are not returned.

## 4.3 Generate skip plan

With the candidate bindings for every variable, we are now ready to generate a plan to compute candidate output tuples. For every sentence where there are candidate bindings, a plan is generated and executed (see the green box in Figure 2).

By now, the set of sentences that are considered are narrowed down to only the sentences of quintuples that are bound to some variables in the previous step. If the extract clause is empty, then all sentences will be considered.

The main idea behind this step is to avoid iterating over variables that have many bindings and instead, rely on the bindings of other variables to determine possible bindings for those variables. This way, the overall number of iterations over all variables is likely to be reduced. For example, a naive implementation of the query in Example 4.1 will require at least 6 nested loops, one for each variable $a$, $b$, $c$, $d$, $v_1$, and $v_2$. where the last two variables were used to represent the respective $\wedge$ in the last condition. Some variables such as $v_1$ is expensive to iterate over as we need to consider all possible spans. Instead, $v_1$ can be easily determined once $a$ and $b$ are determined. Hence, $v_1$ can be skipped initially.

This module (GSP) selects variables to be skipped based on an estimate on the number of bindings they have. To decide which variables to skip initially, GSP exploits the existence of *horizontal conditions*, which are conditions in the extract clause of the form $x = e_1 + e_2 + ... + e_m$ or a condition of the form $(e_1 + ... + e_m)$ eq $(x)$. Here, $x$ is a variable, $m \geq 1$, and $e_i$ is a variable reference, or the subtree of a previously defined variable or $\wedge$, or a path expression. For example, $e = a + \wedge + b + \wedge + c$ of the query in Example 4.1 is a horizontal condition.

---

**Algorithm 2:** Generate Skip Plan

**Input** : normalized query, bindings, sid $s$
**Output**: a list of variables to skip for each horizontal condition

*skip_lists* = empty;
$t$ = number of tokens in sentence $s$;
**foreach** *horizontal condition c in query* **do**
    *cost* = empty;
    **foreach** *variable v in condition c* **do**
        $cost[v] = t(t+1)/2$ if $v$ is $\wedge$. Otherwise,
        $cost[v] = |bindings[v][sid = s]|$

    *sorted_cost* = sort cost in descending order;
    **foreach** $v$ *in sorted_cost* **do**
        $v_l$ = left variable of $v$ in $c$;
        $v_r$ = right variable of $v$ in $c$;
        **if** $v_l$ *and* $v_r$ *are not in skip_lists[c]* **then**
            *skip_lists*[c].append($v$)

**return** *skip_lists*

**Cost model** We adopt a simple cost model where the *cost of evaluating a variable* is proportional to the number of bindings for that variable. We estimate the cost of a variable $x$ as the number of bindings in *bindings[x]*. The *cost of $x$ w.r.t. a sentence with sid $s$* is denoted as $|bindings[x][sid = s]|$, the number of bindings in $|bindings[x]|$ with sentence id equal $s$.

EXAMPLE 4.5. For example, suppose the DPLI module returns the bindings (0,3,-), (1,9,-), (23,5,-), (23,10,-), (35,3,-), ... for the dominant path //verb/dobj//"delicious" of Example 4.1 (we show only the first two components of the quintuples). The estimated cost of evaluating the variable $d = c$/"delicious" for the sentence with sentence id 23 is 2 while the estimated cost for sentence with sentence id 0 is 1. On the other hand, the variable $v_1$ which represents ^ will have an estimated $t(t + 1)/2$ number of spans to consider where $t$ is the number of tokens in the current sentence. □

**A greedy algorithm** As described in Algorithm 2, GSP goes over each horizontal condition of the query to generate a skip plan w.r.t. the current sentence. Based on the estimated cost of each variable, it will greedily select the variable $v$ with the highest cost to be skipped if possible. If the left variable $v_l$ (resp. right variable $v_r$) of $v$ does not exist, then $v_l$ (resp. $v_r$) is assumed not to be in the list of variables to be skipped. A subsequent nested loop over the remaining variables aligns the variables according to the horizontal conditions and evaluates the final results of the query.

Observe that by our definition, ^ are the costliest variables if they exist. We have simplified our analysis of the cost of ^ here but a more refined analysis based on the limits of neighboring tokens of variables can be computed.

EXAMPLE 4.6. For our query in Example 4.1, if the number of tokens in the sentence is sufficiently large, then $v_1$ and $v_2$ will be skipped. So, instead of 6 loops, there will be 4 loops for variables $a$, $b$, $c$, and $d$. The binding for $v_1$ and $v_2$ will be computed based on the values of $a$, $b$, $c$, and $d$. □

**Align skipped variables and check constraints** For a given combination of bindings for variables that are not skipped, we proceed to compute the possible bindings for skipped variables by aligning them to the bindings of the unskipped variables.

EXAMPLE 4.7. We have selected to skip $v_1$ and $v_2$ in Example 4.6. So for each combination of $a$, $b$, $c$, and $d$ values, we compute the bindings for $v_1$ and $v_2$ based on the existing bindings. Since $v_1$ is a span between $a$ and $b$ and $v_2$ is a span between $b$ and $c$, the spans are well-defined by the bindings of $a$, $b$, and $c$ (which is a much easier task than to consider all possible spans for $v_1$ and $v_2$ naively). Finally, we check that the path expressions and constraints among variables, which are generated in Example 4.1, are valid. In other words, we check that $a$ is indeed an entity, $b$ satisfies the path //verb[text="ate"] etc. and we check the six constraints $b$ parentOf $c$, $c$ ancestorOf $d$, $a$ leftOf $v_1$, $v_1$ leftOf $b$, $b$ leftOf $v_2$, $v_2$ leftOf $c$, are satisfied before returning the result as an answer (there is no need to consider satisfying clause for this query). These checks are necessary since, as mentioned earlier, the bindings obtained by evaluating the indices with decomposed paths may still contain false answers. In the event that there are satisfying clauses, such as the query in Example 2.1, the algorithm proceeds to evaluate each clause with the current bindings (see next section). If there is sufficient evidence for every clause, the output tuple is added to the set of results. The next section describes how aggregation is performed. □

## 4.4 Aggregate

For every combination of bindings for *all* variables, the output tuple is returned if the if and satisfying clauses are satisfied, and the excluding clause is not satisfied.

Next, we elaborate on how evidence is aggregated from the text corpus when the satisfying clause is evaluated. While there are several optimizations that can also be performed on this part of the evaluation, we do not elaborate on the optimizations here.

### 4.4.1 Aggregate evidence

Recall that the satisfying clause consists of a set of conditions each with a weight and there is up to one satisfying clause for each output variable. The $i$th condition has weight $w_i$ which corresponds to how important the condition is to determining whether the value in question should be extracted. The score of a value $e$ for the variable under consideration is the weighted sum of confidences, computed as follows:

$$\text{score}(e) = w_1 * m_1(e) + \ldots + w_n * m_n(e)$$

where $w_i$ denotes the weight of the $i$th condition and $m_i(e)$ denotes the degree of confidence for $e$ based on condition $i$. There are different kinds of conditions that KOKO supports to aggregate evidence. The confidence for $e$ is computed for each sentence in the text and aggregated together.

**Boolean conditions** KOKO supports boolean conditions. The first three clauses of the query in Example 2.3 are boolean conditions. The first two conditions can be checked without the text corpus. The condition str($x$) contains ⟨*string*⟩ requires that the string of $x$ contains ⟨*string*⟩, and the condition str($x$) mentions ⟨*string*⟩ requires that ⟨*string*⟩ be a substring of $x$. For example, the string "*chocolate ice cream*" contains "*ice*", mentions "*choc*" but does not contain "*choc*".

A condition $x$ ⟨*string*⟩ (resp. ⟨*string*⟩ $x$) requires that $x$ is strictly followed by (resp. preceded by) ⟨*string*⟩ (e.g., $x$ ", *a cafe*"). In general, regular expressions are supported with the condition "$x$ matches ⟨*pattern*⟩" where ⟨*pattern*⟩ is a regular expression.

For boolean conditions, the degree of confidence $m_i(x)$ is either 0 or 1. Hence, for the query in Example 2.3, as long as $x$ contains "*Cafe*" (resp. "*Roasters*"), we have $m_1(x) = 1$ (resp. $m_2(x) = 1$) regardless of how many times those conditions are satisfied in the text[2]. Similarly, $m_3(x) = 1$ as long as there is a sentence in the corpus where $x$ is followed by "*, a cafe*".

KOKO also supports a more general (and non-Boolean) proximity clause of the form "$x$ near ⟨*string*⟩". For example, for the condition $x$ near "*coffee*", if $x$ = "*Cafe Benz*", then $x$ will score well in the statement "*Cafe Benz serves great coffee*".

A near pattern will generate a matching score inversely proportional to the distance between the mention and the context tokens using the following formula, where distance denotes the number of tokens separating the ⟨*string*⟩ from the candidate $x$.

$$\text{score} = \frac{1}{1+\text{distance}}$$

The excluding clause filters results based on the conditions specified. It supports all of the conditions described above.

**Descriptors: non-Boolean conditions** Descriptor conditions are of the form $x$ [[*descriptor*]] (similarly for left-sided conditions). A descriptor condition "$x$ [[*descriptor*]]" is satisfied if the text after the candidate value provides sufficient evidence of the descriptor. Intuitively, we determine a score for a candidate value w.r.t. a sentence as follows:

(a) the descriptor is first expanded,
(b) the sentence where $x$ occurs in is decomposed, and

---
[2] An alternative semantics is to take the multiplicity of occurrences into account, which we do not consider here.

(c) the similarity scores between the expanded descriptors and decomposed sentences are aggregated.

*(a) Descriptor expansion* A simple expansion strategy is to determine different ways of expressing parts of the descriptor by utilizing word embeddings to replace verb/noun words (e.g., "*serves*", "*coffee*") by similar terms. Using this, we will obtain phrases such as "*sells coffee*", "*host coffee*", "*sells espresso*", etc. However, conventional word embeddings train parameters by optimizing word co-occurrence-based objective functions. Therefore, these embeddings are not directly tuned to represent the semantic relationship between words, such as synonym, hyponym, or hypernym. As a result, this approach could also generate phrases such as "*serves tea*" from the phrase "*serves coffee*", which is not implied by the original descriptor.

KOKO utilizes two techniques to address this issue. First, we employ paraphrase-based word embeddings (e.g., https://github.com/nmrksic/counter-fitting), which uses a paraphrase database as an additional resource to train embeddings. The result is that the embeddings are more likely to express a semantic relationship. Second, we can also employ a simple domain ontology that contains sets of related words (e.g., different coffee drinks such as cappuccino, macchiato). The ontological knowledge allows us to safely replace "*coffee*" by these different coffee drinks.

*(b) Sentence decomposition* A sentence can be long and complex, describing several different aspects simultaneously. As a consequence, even though a sentence may describe the property specified in a descriptor, the descriptor may not match well to the overall sentence due to extraneous noise. To alleviate this problem, we decompose each sentence into a set of canonical sentences, where each canonical sentence is implied by the original sentence.

The work of [2] applies a sentence decomposition technique to generate several canonical clause fragments from each sentence. This decomposition procedure consists of two stages: 1) segment a sentence into canonical clauses, 2) generate shorter fragments by deleting words from each clause. In KOKO, we perform only (1) since our goal is to sum the signals of patterns over individual clauses. Specifically, we applied the sentence decomposition component from the relation triplet extraction systems [42].

*(c) Aggregating the similarity scores between expanded descriptors and decomposed sentences* Let $E(d)$ be the result of expanding the descriptor $d$ in the query. Specifically, let $E(d) = \{(d_1, k_1), ..., (d_m, k_m)\}$ where $d_i$ is an alternate form of $d$ and $k_i$ is a score between 0 and 1 that represents how close is $d_i$ is to $d$.

Let $C = \{(c_1, l_1), ..., (c_n, l_n)\}$ be the set of canonical sentences derived from the original sentence $s$ ($c_i$ is the canonical sentence and $l_i$ is a score). We compute the measure of confidence for the clause $x$ [[$d$]] as the maximum over the aggregate of how well each expanded descriptor matches against each canonical sentence.

conf($x$ [[$d$]]) w.r.t. sentence $s$ = Max$_{i \in \{1,m\}}$(match($d_i, s$)):
match($d_i, s$) = $\Sigma_{j \in \{1,n\}}$(match($d_i, c_j$))
match($d_i, c_j$) = 0 if the word seq. $d_i$ does not occur in $c_j$
match($d_i, c_j$) = $k_i * l_j$ otherwise

Recall that the descriptor $d$ is semantically close to phrases $d_1, ..., d_m$. Hence, the degree by which $d$ matches $s$ is given by the best match, among the expanded descriptors, to $s$. Since $s$ is decomposed into canonical sentences $c_1, ..., c_n$, the extent to which a descriptor $d_i$ matches $s$ is given by the sum of the matches between $d_i$ and each canonical sentence. The overall confidence score of $x$ [[$d$]] w.r.t the entire document is the sum of confidence scores of $x$ [[$d$]] w.r.t every sentence in the document. It remains to describe what is match($d_i, c_j$).

The term match($d_i, c_j$) is 0 if the word sequence of $d_i$ does not occur in $c_j$. A *word sequence* $y_1, ..., y_q$ *occurs in a sentence* $c$ if $c$ contains the words $y_1, ..., y_q$ in this order and each consecutive pair of words $y_i, y_{i+1}$ may be separated by 0 or more words. If $d_i$ occurs in $c_j$, then the confidence score is computed as the product of the respective weights (i.e., $k_i$ and $l_j$). Naturally, other scoring schemes are possible, such as one that takes into account how well a word sequence matches a sentence, preferring contiguous matches over matches with larger gaps between words.

## 5. RELATED WORK

Several information extraction systems are based on training machine learning models or deep learning models to recognize occurrences of typed entities in text (e.g., [16, 30]). However, learning models are typically specific to a particular domain and do not generalize to other domains (e.g., to extract restaurant names instead of cafe names). In addition, learning methods usually require significant annotated training corpus for obtaining a model with sufficient precision and recall. In contrast, KOKO does not require training data and relies on user-defined conditions instead. For many learning models, preparing an annotated corpus is typically the most expensive and time consuming process. Additionally, KOKO is also *debuggable* where users can discover the reasons that led to an extraction, which is important for certain applications [13].

Rule-based systems have traditionally used cascades of finite state automatas (e.g., [1, 3, 17]) and systems have been developed using the Common Pattern Specification Language (CPSL) (e.g., [10, 43]). Odin [44] also uses CPSL style grammar and allows one to specify patterns over dependency trees. However, it inherits the issues that were described in [12], including lossy sequencing and rigid matching priority. Semgrex [38] also supports querying over dependency trees. However, 1) Semgrex is not declarative and works only in the procedural environment of Java, 2) Semgrex neither supports surface-level patterns nor aggregation over evidences, and 3) Semgrex does not exploit any indexing techniques. Xlog [39] developed a datalog-like language for information extraction, where users can define extraction predicates to be "plugged in" the datalog-style rules. Xlog includes a number of query optimization techniques centered around how to execute thousands of such rules over a document. SystemT/AQL [12] is a powerful commercial rule-based language that allows users to specify SQL-like views over text through extraction statements that use regular expressions and built-in predicates. Such views can be composed to form a powerful mechanism for writing modular and composable extraction tasks. IKE [18] is another system that extracts entities from text via a language based on patterns over the syntax and POS tags of sentences. In addition, IKE supports distributional similarity based search. For example, "dog $\sim$ 20" would find 20 words similar to "*dog*" and match based on "*dog*" and the similar words. Hence, a KOKO descriptor such as "$x$ [[serves coffee]]", where $x$ is an entity, has a similar effect to writing "NP ([[serves coffee]] $\sim$ 20)" in IKE, where NP refers to a noun phrase. (KOKO descriptors now default to a fixed number of expanded terms. We plan to allow the user to select among the expansions in future.) However, all systems (Xlog, SystemT/AQL, IKE) do not support querying over dependency structures. While there are other differences across all these systems, KOKO differentiates itself from existing systems in its ability to combine conditions on the surface text, conditions on the dependency trees and to support conditions that are robust to linguistic variations.

A related approach is to use patterns [6, 24] to perform information extraction. Some techniques seed and learn patterns to extract the desired information (e.g., [15, 23, 35, 36] to name a few). For this method to work well, the seed words must typically be "representative" of the desired set of words to be extracted. A notable system in this general category is NELL, which is a system that constantly extracts structured information from the Web, and uses its acquired knowledge to find new rules for extracting new information [8, 29]. NELL is able to learn the relationship between entities as well as the categories each entity belongs to (e.g., "*Starbucks*" is a cafe). Thus, it can potentially be used to find entities of a certain type (e.g., cafes). However, there are two fundamental differences between how NELL and KOKO perform extraction: (1) NELL requires a few seed examples to learn about a new category, while KOKO relies on the human intelligence that is embedded in the query it runs, and (2) KOKO queries are tailored to deal with a specific corpus while NELL reads the Web. In other words, a KOKO query that performs well on one type of corpus may not perform as well on random documents from the web. On the other hand, NELL is more conservative and only relies on patterns that are robust. Consequently, the results from NELL generally have a high precision but lower recall (compared to other systems that are tailored for specific extraction tasks). Our experiments with NELL (see section 6.1) further confirms this observation.

Another line of work on querying text queries over *constituency-based parse trees* [27] and is targeted at extracting a single span of tokens as opposed to a tuple of spans of different types. We choose to design KOKO over dependency trees instead due to the availability of fast and accurate dependency parsers today, some of which are open-source [11, 22, 40].

In the next section, we discuss more related work on indexing schemes that have been developed to query linguistic parse trees which we used for comparison in our experimental evaluation.

## 6. EXPERIMENTS

We evaluate KOKO with both real-word and synthetic queries on publicly available text corpora. Our examples throughout the paper (and additional ones in Section 6.3) demonstrate the power of querying both the surface text and dependency tree. Here we experimentally evaluate the usefulness of KOKO's ability to aggregate evidence from different parts of the text in order to capture the right results. We also show that the KOKO engine is efficient and scalable in evaluating KOKO queries. It outperforms alternative indexing techniques and the baseline evaluation algorithm.

KOKO is implemented in Python with spaCy [40] as our dependency parser. We use PostgreSQL as our backend to store our indices and text corpora. Our experiments were executed on a 64-bit machine with 122GB RAM and a 2.3 GHz processor. We use 122GB RAM only to speed up the index creation time for all methods we compare. Our index construction also works with a smaller amount of memory (e.g., 16GB).

### 6.1 Usefulness and quality of extraction

In this section, we demonstrate the usefulness of the KOKO language and the quality of its extraction with the task of extracting entities with relatively rare mentions.

Specifically, we focus on using the conditions supported in the satisfying clause to aggregate evidence and extract the names of new and upcoming cafes from blog posts which we scraped from two well-known cafe websites, BARISTAMAG [5] and SPRUDGE [41]. We built a ground-truth for the BARISTAMAG and SPRUDGE datasets by crowdsourcing the task of annotating cafes in the corpus on CrowdFlower. Each cafe blog was shown to five workers and we selected any entity with 3 or more votes as part of the ground truth. We ended up with 84 articles and 137 labeled cafe names in the BARISTAMAG dataset and 1645 articles and 671 cafe names in SPRUDGE. We compare the performance of KOKO against IKE [18],

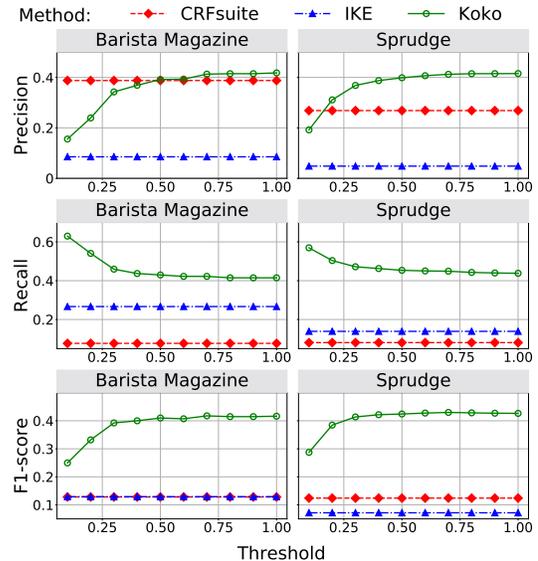

Figure 3: Extracting cafe names with IKE, CRFsuite and KOKO

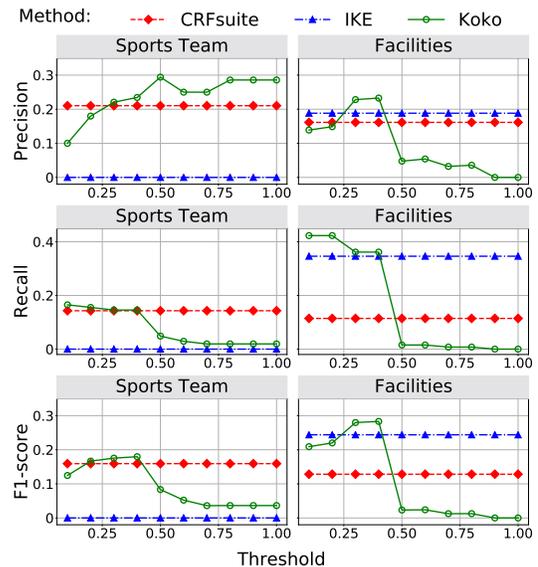

Figure 4: Extracting sports teams and facilities from tweets with IKE, CRFsuite and KOKO

CRFsuite [31], and NELL [8, 29]. We compare to IKE because IKE supports distributional similarity based search which is similar to KOKO descriptors as we described in Section 5. The other baseline, CRFSuite, is an open-source implementation of Conditional Random Fields (CRF) – a popular machine-learning technique for the Named Entity Recognition (NER) task. We used the averaged perceptron algorithm to train a first order Markov CRF. The features used in the CRF model include the tokens along with their preceding and following tokens, prefix and suffix of each token up to 3 characters, and set of binary features that test if token matches a few regular expressions (mostly to test if it has digits, or if the token is all digits and so on). Note that the CRF baseline is included as a representative of machine learning techniques to demonstrate that such techniques require a considerable amount of training data to achieve reasonable performance. Of course, more complex models such as neural networks can be used but they require even more training data. Finally, we compare KOKO with NELL.

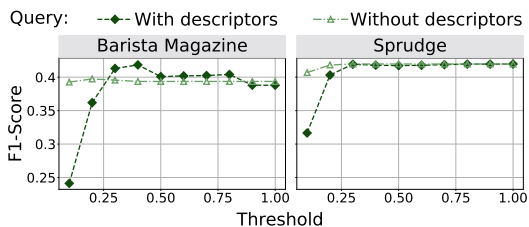

Figure 5: KOKO with/without descriptors

**Writing queries:** We followed a simple approach to write a query for extracting cafe names. (1) examine a few positive examples from the corpora and studying the frequent patterns, (2) write conditions that capture both the observed patterns as well as our background knowledge on the entities, and (3) examine the results obtained by the written query and adding conditions to exclude the frequent mistakes. We divided the patterns in the satisfying clause into 3 categories: high, medium and low confidence, and gave weights of 0.8, 0.5, and 0.2 respectively. It is important to note that the patterns we learn from step (1) are independent of the system we use to discover the entities. Thus, our goal in steps (2) and (3) is to capture these patterns to the extent possible using the features of KOKO or IKE. In Appendix A, we list the query for extracting cafe names and queries for extracting other entities.

**Extraction results:** Figure 3 shows the precision, recall and F1-score achieved by the three systems on the task of extracting cafe names. The x-axis shows the threshold value used in the satisfying clause of the KOKO query. The values reported for IKE and CRFsuite appear as horizontal lines as they do not depend on the thresholds used in the KOKO query. KOKO performs better than IKE and CRFsuite (in terms of F1-score) for all thresholds for both datasets with its best performance achieved at 0.6. The main reason is due to KOKO's ability to aggregate partial evidence from *multiple* mentions in the document. In contrast, IKE only considers single sentences and cannot aggregate partial evidence. Similarly, CRFsuite, by design, focuses on single mentions of an entity and judges if it should be extracted regardless of other mentions. Note that we used 50% of the available data to train the CRFsuite algorithm. Our experiments show that our results with KOKO outperform CRFsuite even when up to 90% of the data is used for training.

**Capturing linguistic variations:** We also studied how much the descriptor operator contributes to the extraction of desired entities. To do so, we executed our KOKO query for extracting cafe names with and without descriptor expansions. The use of descriptors in BARISTAMAG yields better results while no improvement can be observed for SPRUDGE. See Figure 5. This is mainly because the BARISTAMAG articles are generally shorter than SPRUDGE articles (480 vs. 760 words per article). For shorter articles, KOKO needs to rely on weak signals obtained from descriptors, but in the presence of longer text using descriptors are not as effective.

**NELL for entity extraction [8, 29]:** Nell is a system for extracting structured information from the Web. The creators of NELL kindly accepted our request and added "cafes" as a new category in NELL (on Feb. 1, 2017) with 17 seed instances. Within a month NELL discovered 72 patterns for extracting cafe names.

BARISTAMAG: precision=0.7, recall=0.05, F1=0.1
SPRUDGE: precision=0.27, recall=0.04, F1=0.06

The reason for the low recall is that NELL works well on cafes (or entities, in general) that are mentioned frequently on the Web, while the cafes we extract occur only a few times.

**Other entity types:** Here, we present our additional experiments on extracting different entity-types, namely sport teams and facilities. The obtained results are aligned with our observations for the task of extracting cafe name, but they are presented here for completeness. The figure below shows the precision, recall and the F1-score for the task of extracting sports teams and facilities from the WNUT dataset (a collection of tweets in which different types of named entities are labeled). Note that KOKO outperforms the baselines (in terms of F1-score) in both cases when the threshold is set to 0.4 (Recall that the x-axis is the threshold value of KOKOs with threshold construct). An interesting observation is that the performance of baselines is much closer to KOKO compared to the task of extracting cafe names. This is due to the fact that tweets are very short and stand-alone documents, and KOKO's ability to compile evidence from multiple parts of the text cannot be exploited.

## 6.2 Performance of KOKO modules

Our performance experiments make use of two text corpora:

| corpus | #sentences | size | notes |
| --- | --- | --- | --- |
| HappyDB [4] | 140K | 13MB | $\sim$ 100K happy moments |
| Wikipedia [45] | 110M | 8.9GB | $\sim$ 5M articles |

### 6.2.1 Index construction time and size

We compare the construction time of KOKO's indices with one baseline algorithm and two prior indexing techniques. We also use these methods to compare the performance of our DPIL module before we evaluate the effectiveness of our ability to aggregate evidence and KOKO's overall performance. For each of the indexing techniques, we create the necessary indices (B-tree) in PostgreSQL.

**INVERTED:** This is our baseline indexing technique where the index maps each label to the sentence id and token id pairs that contain that label. We store this index as a table with the following schema: *P(label, sentence id, token id)*. Given a KOKO query, we retrieve from the table all sentences that contain all labels in the query with one nested-SQL query.

**ADVINVERTED [7, 20]:** Bird *et al.* developed an advanced inverted indexing technique to query the constituency-based linguistic parse trees with their LPath queries. This indexing technique depends on a labeled form of linguistic trees that is stored in a database with the schema: *P(label, sentence id, token id, left, right, depth, pid)*, where *pid* is the id of the parent node in the tree. This schema allows us to express various relationships between nodes in linguistic trees. For example, to specify that *c* is a child of *p*, we write: *c.sentence id = p.sentence id* and *c.pid = p.token id*. Compared to INVERTED, ADVINVERTED allows more precise expression over KOKO queries. Given a KOKO query, we also translate the path expressions into a nested-SQL query to retrieve, from the table, all sentences that passed the conditions of the extract clause of the KOKO query.

**SUBTREE [14]:** Instead of individual labels, Chubak and Rafiei's SUBTREE index stores every unique subtree, up to a maximum size, as index keys. Given a tree-structured query, they decompose the query into multiple subtrees that are all no larger than the maximum size, and then search through the index to find sentences that contain all the decomposed subtrees and furthermore, contain the same structure as the query. This indexing technique is primarily designed for indexing the linguistic parse trees, which tend to have a reasonably small number of subtrees because of their small branching factor.

For our experiments, we implemented [14] with *mss* = 3 (mss is the maximum subtree size for indexing) and the *root-split* coding schema. Note that a fundamental difference between KOKO and SUBTREE index is that the latter focuses on indexing constituency-based parse trees which has only a single type of labels. Hence, for our experiments, we have to create two SUBTREE indices, one for parse labels and one for POS tags. We join the root nodes of subtrees returned from the two indices when needed. However, this operation may hurt the *index effectiveness* (defined later), which corresponds to how precise the index is in returning the sentences that will pass the conditions of the extract clause, since joining the root nodes does not guarantee that the two subtrees are referring to the same set of tokens. In addition, SUBTREE index with root-split coding schema does not support wildcards. With this implementation, SUBTREE index supports 125 queries (out of 350) in our Synthetic Tree benchmark (to be described shortly).

**KOKO:** As described earlier, we use a multi-indexing scheme: inverted word (W) and entity (E) indices, hierarchy PL and POS indices. $W$ and $E$ are stored as follows:

$W(word, \underline{x, y, u, v, d}, pl_{id}, pos_{id})$, $E(entity, \underline{x, u, v})$

In table $W$, the underlined $x, y, u, v, d$ attributes correspond to the quintuple $(x,y,u\text{-}v,d)$. In addition to the quintuple, we also store the node id ($pl_{id}$) of the word in the parse label (PL) index and node id ($pos_{id}$) of the word in the Part-Of-Speech (POS) index. As we shall describe shortly, the ids ($pl_{id}$ and $pos_{id}$) are used to access the posting lists of the PL and POS index. In table $E$, the underlined $x, u, v$ attributes map to the triple $(x, u, v)$.

We use two additional tables, *PL* and *POS*, with the same schema to store the *Closure Table* [25] of the nodes in the hierarchy index for parse labels and POS tags respectively:

*PL/POS(id, label, depth, aid, alabel, adepth)*

where *id, label*, and *depth* represent the current node id, node label, and node depth respectively and *aid, alabel*, and *adepth* stand for the ancestor node id, label, and depth respectively. We retrieve the posting list of the PL index (resp. POS index) by joining *PL* table (resp. *POS* table) with table $W$ over attribute *PL.id* (resp. *POS.id*) and attribute $W.pl_{id}$ (resp. $W.pos_{id}$). Given a KOKO query, we translate it into a nested-SQL query composed by sub-queries that search for the matching nodes from the *PL* and *POS* tables [9, 25], and conditions that joins with the $W$ and $E$ tables.

**Performance** Figure 6(a) shows the index construction time for the Wikipedia dataset as we vary the number of articles using the four indexing techniques described above. INVERTED and ADVINVERTED require similar but less time to construct than KOKO since they need not construct and store the hierarchy index for parse labels and POS tags as in KOKO. KOKO is more than $2\times$ faster compared to SUBTREE, where significant amount of time is spent on enumerating every unique subtree (with at most 3 labels) for sentences in the input text corpus. In contrast, KOKO enumerates over the parse tree only once for each sentence and for each type of label. Since the HappyDB corpus is smaller, it requires about $10+$ minutes to construct for ADVINVERTED, INVERTED, and KOKO. For SUBTREE, the construction time is about $30+$ minutes.

Figure 6(b) shows the amount of disk space, including the posting list and all necessary indices in PostgresSQL, used by the four indexing techniques for the Wikipedia corpus. KOKO uses the least space due to the compact nature of hierarchy indices in merging parse trees of different sentences. INVERTED uses a slightly less space than ADVINVERTED since it does not store the leftmost, rightmost, and parent information for each label. On the other hand, SUBTREE uses the most space as it stores all the unique subtrees in the text corpus, which adds up to a few times more than the size of the original corpus. For HappyDB, a similar phenomenon occurs.

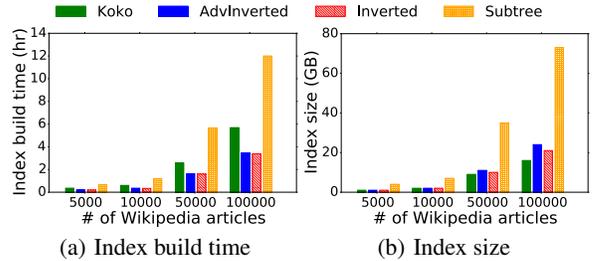

(a) Index build time  (b) Index size

Figure 6: Index construction with increasing size of input text corpus. KOKO takes longer time to build than INVERTED and ADVINVERTED, but has the smallest footprint.

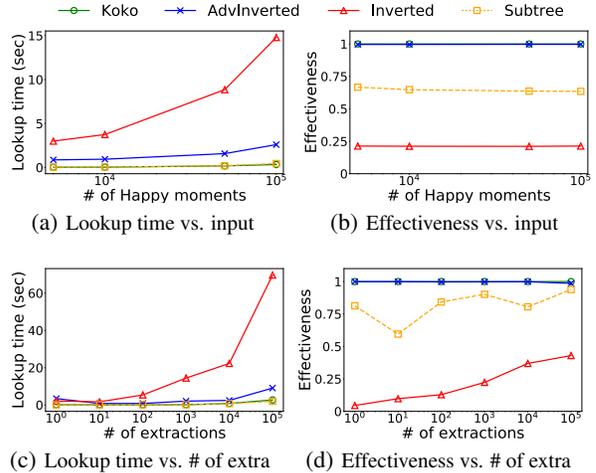

(a) Lookup time vs. input  (b) Effectiveness vs. input

(c) Lookup time vs. # of extra  (d) Effectiveness vs. # of extra

Figure 7: Index performance on HappyDB dataset. KOKO is more efficient and effective than the other indexing techniques. Note that SUBTREE supports a subset of our benchmark queries that do not contain word attribute and wildcards.

KOKO requires the least space, using 0.67GB. Both INVERTED and ADVINVERTED use 1.1GB and SUBTREE uses 3.2GB of space.

### 6.2.2 Performance of DPLI module

We evaluate the performance of the indexing techniques on HappyDB and Wikipedia with two metrics: (1) The index lookup time (or lookup time for short); and (2) the *index effectiveness score*, which is the ratio of the number of sentences that contain the bindings for all variables in the given query to the number of sentences returned by the index.

**Synthetic Tree benchmark** To thoroughly evaluate the performance of the indexing technique in our DPLI module, we generate a benchmark with 350 synthetic queries with node variables that is defined to form paths or tree patterns. We first generate a set of queries with a single node variable that is defined by a path. We then create other queries by modifying the length of the path from 2 to 5, the attribute types along the path (parse labels, parse labels + POS tags, parse labels + POS tags + text), whether the query contains a wildcard or not, and whether it starts from the root level or not. For example, the variable definition, $v$ = /root/punct/advmod, defines a variable $v$ with 3 levels. It contains only *"parse label"* type, has *no wildcard*, and *starts from the root level*. Under each setting, we generate 5 random queries with different selectivity. We then generate queries with multiple node variables that form a tree and we vary the number of labels in the tree pattern from 3 to 10. For example, variables, $x$ = /root/punct/advmod, $c_1$ = $x$/advmod, and $c_2$ = $x$/det, form a tree pattern with 5 labels. For each set-ting, we generate 5 random queries with different selectivity. The benchmark can be downloaded[3].

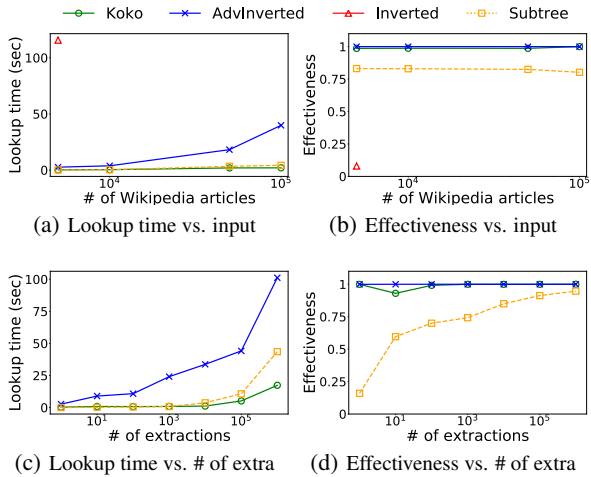

(a) Lookup time vs. input  (b) Effectiveness vs. input
(c) Lookup time vs. # of extra  (d) Effectiveness vs. # of extra

Figure 8: Index performance on Wikipedia dataset. KOKO remains highly efficient and effective over large-text corpora, e.g., 100K articles with 2.2M sentences. INVERTED does not scale over 5K articles and performs badly even for small-text corpora.

**Performance**   Figure 7(a) (resp. Figure 8(a)) shows the index lookup time with increasing number of sentences (resp. number of articles) in the text corpus: KOKO and SUBTREE are much faster than INVERTED and ADVINVERTED because the indices of the former directly accounts for the hierarchical structure of the parse trees while the latter considers only the labels and is largely agnostic to the structure. Even though ADVINVERTED uses additional attributes (*left, right, depth, and parent*) to store structural information, validation over the hierarchical conditions requires additional computation and takes longer time to execute in general. INVERTED does not consider any hierarchical conditions in the queries and thus often results in significantly larger intermediate results. It requires significantly longer time to execute in general and fails to scale over 5000 articles on Wikipedia text corpora. Figure 7(c) and Figure 8(c) show index lookup time over queries with increasing number of extractions (i.e., tuples returned). This is expected since the sizes the intermediate results tend to be higher for queries with larger number of extractions. Consequently, the cost of joining over these intermediate results will be larger. Among all techniques, KOKO scales the best over the number of matches. This is because by leveraging the hierarchy index, KOKO is able to find the matching nodes for parse labels and POS tags without joining over the post list. SUBTREE performs better than ADVINVERTED as it contains less number of joins over the post lists in general.

Figure 7(b) and (d) and similarly Figure 8(b) and (d) demonstrate the effectiveness of the indexing techniques on HappyDB and Wikipedia. As shown, KOKO and ADVINVERTED perform much better than the other two techniques, obtaining near perfect effectiveness rate over all different settings. SUBTREE achieves above 0.6 effective rate, however, due to the limited information (only the root node) that it stores, it performs less accurately for queries with more than one output attribute. INVERTED dismisses the hierarchical conditions in the queries, and thus performs badly (with less than 0.5 effectiveness rate) over all settings on both text corpora.

### 6.2.3  Performance of GSP module

[3]Both SyntheticTree and SyntheticSpan (described in Section 6.2.3) benchmarks can be downloaded at: *https://github.com/rit-git/koko_benchmark*.

|  | HappyDB |  |  | Wikipedia |  |  |
|---|---|---|---|---|---|---|
| # of atoms | 1 | 3 | 5 | 1 | 3 | 5 |
| KOKO&GSP | 0.28 | 0.35 | 0.37 | 0.19 | 0.28 | 0.36 |
| KOKO&NOGSP | 0.19 | 1.27 | 290.42 | 0.17 | 5.76 | 607.48 |

Table 1: Average evaluation time (ms/per sentence) over extract clause on sentences in HappyDB and Wikipedia.

**Synthetic Span benchmark**   We generated another benchmark, Synthetic Span, to evaluate KOKO's Generate Skip Plan (GSP) module. Recall that GSP is designed to improve the performance of KOKO while evaluating variables with multiple atoms in a horizontal condition, e.g., $x = e_1 + e_2 + e_3$. Therefore, in our Synthetic Span benchmark, we included span variables defined with 1, 3, and 5 atoms with at most 0, 1, and 2 atoms to skip (at most) respectively. For example, the span variable, $v$ = //verb + ^ + /root/xcomp + ^ + "*happy*", consists of 5 atoms: //verb as a verb, ^ as a span of arbitrary length, /root/xcomp as a token where root and xcomp are parse labels, and "*happy*" as single word. Under each setting, we randomly generate 100 queries with varying selectivity. In total, SyntheticSpan benchmark contains 300 queries.

**Performance**   We further evaluate the GSP module and demonstrate that the GSP module is effective in enhancing the performance of KOKO. Here, we compare KOKO with GSP (KOKO&GSP) with a baseline approach (KOKO&NOGSP), which uses nested-loops to evaluate every variable in a query according the order of their definitions. We use HappyDB and the Wikipedia dataset and our Synthetic Span benchmark for the evaluation.

As shown in Table 1, KOKO&NOGSP performs slightly better than KOKO&GSP when there is only one atom in the span variable. This is because the cost of skip plan generation outweighs the cost of savings obtained by skipping some of the variables. However, when there are more atoms, KOKO&GSP is more than three orders of magnitude faster than KOKO&NOGSP.

## 6.3   Scalability and overall performance

We now evaluate how KOKO performs when given a large input text corpus on the three queries for different extraction tasks.

**Chocolate** (Low):   extract $c$:Entity from wiki.article if (
*A heuristic query*   /ROOT:{
*for chocolate types*   $v$ = //verb, $o$ = v/pobj[text="chocolate"],
  $s$ = v/nsubj } (s) in (c))
  satisfying $v$
  (str($v$) ~ "is" {*1*})
**E.g.,** *Baking chocolate (c) is a type of chocolate that is prepared or manufactured for baking.*

**Title** (Medium):   extract $a$:Person, $b$:Str from wiki.article if (
*A heuristic query*   /ROOT:{
*for people's titles*   $v$ = //"called", $p$ = v/propn, $b$ = $p.subtree$,
  $c$ = $a + ^ + v + ^ + b$})
**E.g.,** *Cyd Charisse (a) had been called Sid (b) for years.*

**DateOfBirth** (High):   extract $a$:Person, $b$:Date from wiki.article if (
*A heuristic query for*   /ROOT:{$v$ = verb})
*people and their DOB*   satisfying $v$
  (str($v$) ~ "born" {*1*})
**E.g.,** *He was married to Alys Adle Thomas on 1 December 1900 in London, and the couple had a daughter Vera Alys (a) born in 1911 (b).*

We measure the *selectivity of a query* as the ratio of the number of articles that contain one or more extractions to the total number of articles. Queries with more variables tend to be more selective. The DateOfBirth query has only 3 variables so that it will match with a high percentage of sentences. The first two queries have 4 and 5 variables respectively and their selectivities are, respectively, low ($< 1\%$), medium ($\sim 10\%$), and high ($> 70\%$).

|   | Size† | Normalize | DPLI | LoadArticle | GSP | extract | satisfying |
|---|---|---|---|---|---|---|---|
| C | 5K | 0.05 | 0.79 | 2.67 | 0.00 | 0.04 | 0.15 |
|   | 50K | 0.05 | 5.58 | 5.71 | 0.00 | 0.16 | 0.86 |
|   | 500K | 0.05 | 58.62 | 72.24 | 0.07 | 1.89 | 9.72 |
|   | 5M | 0.05 | 518.13 | 486.31 | 0.48 | 11.50 | 64.46 |
| T | 5K | 0.015 | 1.54 | 42.51 | 0.02 | 0.65 | 0 |
|   | 50K | 0.015 | 12.33 | 279.31 | 0.19 | 5.33 | 0 |
|   | 500K | 0.015 | 127.41 | 3018.52 | 2.05 | 61.49 | 0 |
|   | 5M | 0.015 | 1350.77 | 15325.64 | 22.77 | 489.28 | 0 |
| D | 5K | 0.028 | 0.61 | 117.21 | 0.60 | 25.09 | 32.36 |
|   | 50K | 0.028 | 4.38 | 984.38 | 5.30 | 227.17 | 322.72 |
|   | 500K | 0.028 | 45.33 | 10843.30 | 55.92 | 3054.06 | 3324.68 |
|   | 5M | 0.028 | 411.69 | 68945.81 | 328.25 | 17949.82 | 24526.23 |

Table 2: KOKO execution time (sec) for three example queries with increasing text corpora. Size† is the # of articles in the text corpora; LoadArticle is the time for loading the candidate articles returned by KOKO index from the DBMS.

**Koko** Table 2 shows the execution time of KOKO as we increase the number of articles of Wikipedia, from $5K$ to $5M$. The total execution time of KOKO is linear in the number of articles in the text corpora and queries with higher selectivity often require longer time to evaluate as there are larger number of extractions.

We further analyze the breakdown of time KOKO spends on different components for each of the queries. Note that this distribution is highly dependent on the complexity and selectivity of the query. As shown, among all queries on different text corpora sizes, KOKO spends less than $2\%$ of its total execution time for normalizing the query (Normalize) and generating the Skip Plan (GSP). The ratio of time KOKO spends on index lookup (DPLI) heavily depends on the selectivity of the queries: queries with higher selectivity requires less percentage of time for index lookup. For example, the "DateOfBirth" query requires only $2\%$ of the time for index lookup whereas "Chocolate" query requires $20\% - 40\%$. Note that the actual time queries spent on index lookup is different with this percentage: queries with higher complexity, e.g., more variables, and higher selectivity often require longer index lookup time. Before evaluating the extract clause and satisfying clause, KOKO will load the parsed articles (LoadArticle) that contain the sentences of interest, as returned by the index lookup, from DBMS into memory. Loading such articles requires a more than $50\%$ of the time for all example queries. The percentage of time KOKO spends for evaluating extract clause and satisfying clause also depends on the complexity of the query and the selectivity of the query: queries with fewer constraints in the extract clause and satisfying clause require less percentage of time to evaluate, e.g., the "Title" query with no constraints in satisfying clause only requires $2\%$ for evaluating the two clauses; In addition, queries with lower selectivity normally require less percentage of time than queries with higher selectivity, e.g., the "Chocolate" query requires $\sim 6\%$ of the time for evaluating the extract and satisfying clauses whereas the "DateOfBirth" query requires more than $30\%$.

**Odin [44]** We also ran Odin (see Section 5) by translating the three queries above to Odin's syntax to the extent possible. Since Odin does not aggregate evidence, our translated queries contain only extract clauses. Furthermore, since Odin iteratively evaluates all patterns until no further matches are found, we helped Odin by specifying the priorities of the patterns which speeds up the execution. We compared the execution time of KOKO (which includes the time to execute the satisfying clause) and Odin for all three queries using 5000 documents from Wikipedia. Odin is 40x, 23x, and 1.3x slower for "Chocolate", "Title", and "DateOfBirth" queries respectively. On 50000 documents, Odin took more than 2 days to complete the annotation and execution of all queries.

# 7. CONCLUSION AND FUTURE WORK

We described KOKO, an extraction system that leverages new NLP techniques and scales nicely with novel indexing and query processing techniques.

Since KOKO queries may use dependency trees, the quality of the dependency parsers impacts the quality of extractions. However, KOKO's ability to aggregate evidence before yielding an extraction can counteract some of the errors introduced by the parsers.

KOKO now relies on the user to specify the dependency patterns, conditions, and weights of queries. A challenge is to design an effective user interface for KOKO to alleviate the burden of specifying a query from the user. and exploit machine-learning models to automatically suggest patterns and weights based on examples. There are further optimization opportunities for KOKO, including adding optimizations steps and parallelizing the evaluation of satisfying clauses, developing techniques (like in [39]) centered around executing a large number of queries/patterns over a document. One can also add to the expressiveness of KOKO. For example, by not limiting to only disjunctions of conditions in the satisfying clauses. More interestingly, we would like to add to KOKO the capability to query over frames or predicates [19, 32, 37] that are evoked by sentences, thus adding yet another layer of semantic-based querying on top of its existing capabilities. A preliminary version of KOKO is already publicly available at [26] and we plan to open-source the complete KOKO system in the near future.

# APPENDIX

## A. QUERIES FOR ENTITY EXTRACTIONS

### A.1 Extracting cafe names

*Koko queries*

Figure 9 demonstrates the query we constructed to extract cafe names from cafe blogs. Using a few blogs, we have reviewed and studied some positive examples. We noticed that a good number of cafes have specific words appearing in their title (lines 3-5). Also, we observed that most blogs are written to introduce a cafe or review what they sell, thus there are some phrases that are likely to appear in the text (lines 6-11). At this point, the query captures most positive examples that we observed in the sampled blogs, but we can improve our query by writing clauses that capture our background knowledge on cafes. For instance, we know that cafes have baristas, a coffee menu, and so on. Lines 12-19 captures these insights.

We then executed the query and immediately notices common incorrect entities that the query is extracting. The most common mistake was extracting locations and street addresses as cafes. This is due to the fact that there are many sentences of the following nature in the corpus:
- *Portland produces and sells the best coffee.*
- *The new cafe on Mission St. has the best cup of espresso.* To avoid such mistakes, we have added the appropriate exclude clauses (lines 34-38). We also observed other types of errors (listed below) which we handled with other exclude clauses:

- Extracting the name of coffee festivals and contests (excluded in lines 25-27)
- Extracting coffee-shop's domain names (excluded in line 23)
- Extracting brand names that are associated with coffee and coffee machines (excluded in lines 29-33)

Lastly, we need to mention how we assigned a weight to each clause. We followed a simple strategy to assign the weights: For any pattern that we are certain to be indicative of a cafe name, we set the weight to 1 to guarantee that the matched tokens are extracted; We then divided the other clauses into *more-likely* and *less-likely* groups and assigned a weight value of 0.02 and 0.015 to them respectively [4].

*IKE queries*

Similar to KOKO, IKE also requires a query to execute. To write IKE queries, we followed the same steps that we discussed earlier.

That is, we reviewed positive samples, identified the common patterns, and implemented conditions that best match our insights using the expressive power of the system. Specifically, for the query in Figure 9, we created first a relation to store cafe names and ran IKE as follows:

The clauses of Lines 3-5 cannot be expressed in IKE and are hence omitted. We express lines 6-7 as follows:
```
"cafe called" (NP)
"cafes such as" (NP)
```
NP denotes noun phrase. It is a built-in POS tag pattern from IKE. In other words, it extracts only noun phrases that satisfy the conditions stated. Each line is executed separately in IKE over the cafe blogs separately and the results are incrementally added to the relation we created.

Line 8 cannot be expressed in IKE and is hence omitted. Lines 9-19 are executed as the following patterns in IKE:
```
(NP) ("sells coffee" ∼ 10)
(NP) ("serves coffee" ∼ 10)
("coffee from" ∼ 10) (NP)
("baristas of" ∼ 10) (NP)
(NP) ("baristas" ∼ 10)
(NP) ("barista champion" ∼ 10)
("barista champion" ∼ 10) (NP)
(NP) ("pour-over" ∼ 10)
(NP) ("french press" ∼ 10)
(NP) ("coffee menu" ∼ 10)
("coffee menu" ∼ 10) (NP)
```
Like before, each line is executed separately on IKE over the cafe blogs and the results are incrementally added to a relation. We omitted the excluding clause as it cannot be expressed in IKE.

### A.2 Extracting sports teams and facilities

Figure 10 and 11 show the koko queries we have used to extract facilities and sports teams from the collection of tweets in the WNUT dataset. Note that the weights we have assigned to each clause is much higher compared to the Koko query we have written for extracting cafe names from blogs. This is because tweets are extremely short documents and there is not risk of exceeding the overall score of 1.

---
[4] Note that only the relative values of weights are important (i.e., assigning 0.2 and 0.15 as weights would produce the same relative scores for entities). We selected smaller weights to make sure that total score accumulated from all clauses wouldn't exceed 1 which is the maximum possible score in our system.

```
 1: extract x:Entity from <InputFile> if ()
 2: satisfying x
 3:     (str(x) contains "Cafe" {1}) or
 4:     (str(x) contains "Café" {1}) or
 5:     (str(x) contains "Coffee" {1}) or
 6:     ("cafe called" x {1}) or
 7:     ("cafes such as" x {1}) or
 8:     (x near ", a cafe" {1}) or
 9:     (x [["sells coffee"]] {0.02}) or
10:     (x [["serves coffee"]] {0.02}) or
11:     ([["coffee from"]] x {0.015}) or
12:     ([["baristas of"]] x {0.015}) or
13:     (x [["baristas"]] {0.015}) or
14:     (x [["barista champion"]] {0.015}) or
15:     ([["barista champion"]] x {0.015}) or
16:     (x [["pour-over"]] {0.015}) or
17:     (x [["french press"]] {0.015}) or
18:     (x [["coffee menu"]] {0.015}) or
19:     ([["coffee menu"]] x {0.015})
20: with threshold τ
21: excluding
22:     (str(x) matches "[a-z 0-9.]+") or
23:     (str(x) matches "@[A-Za-z 0-9.]+") or
24:     (str(x) matches "[Cc]offee|[Cc]afe|[Cc]afé") or
25:     (str(x) matches "[A-Za-z 0-9.]*[Bb]arista [Cc]hampionship") or
26:     (str(x) matches "[A-Za-z 0-9.]*[Bb]rewers [Cc]up") or
27:     (str(x) matches "[A-Za-z 0-9.]*[Ff]est(ival)?") or
28:     (str(x) matches "Coffee News") or
29:     (str(x) matches "[Ll]a Marzocco") or
30:     (str(x) matches "[Ss]ynesso") or
31:     (str(x) matches "[Aa]eropress") or
32:     (str(x) matches "[Vv]60") or
33:     (str(x) matches "CEO") or
34:     (str(x) matches "[0-9]+ [0-9A-Z a-z]+ [Ss]t.?") or
35:     (str(x) matches "[0-9]+ [0-9A-Z a-z]+ [Ss]treet") or
36:     (str(x) matches "[0-9]+ [0-9A-Z a-z]+ [Aa]ve.?") or
37:     (str(x) matches "[0-9]+ [0-9A-Z a-z]+ [Aa]v.?") or
38:     (str(x) matches "[0-9]+ [0-9A-Z a-z]+ [Aa]venue") or
39:     (str(x) in dict("Location"))
```

Figure 9: The query for extracting cafe names

```
 1: extract x:Entity from <InputFile> if ()
 2: satisfying x
 3:     ("at" x {1}) or
 4:     ([["went to"]] x {0.8}) or
 5:     ([["go to"]] x {0.8})
 6: with threshold τ
 7: excluding
 8:     (str(x) contains "p.m.") or
 9:     (str(x) contains "a.m.") or
10:     (str(x) contains "pm") or
11:     (str(x) contains "am") or
12:     (str(x) mentions "@") or
13:     (str(x) contains "today") or
14:     (str(x) contains "tomorrow") or
15:     (str(x) contains "tonight")
```

Figure 10: The query for extracting facilities

```
1: extract x:Entity from <InputFile> if ()
2: satisfying x
3:     (x [["to host"]] {0.9}) or
4:     (x "vs" {0.9}) or
5:     ("vs" x {0.9}) or
6:     (x "versus" {0.9}) or
7:     (x [["soccer"]] {0.9}) or
8:     ("go" x {0.9})
9: with threshold τ
```

Figure 11: The query for extracting sports teams